# Reversible and nonvolatile manipulation of the spin-orbit interaction in ferroelectric field-effect transistors based on a two-dimensional bismuth oxychalcogenide

Ming-Yuan Yan,[1,3] Shuang-Shuang Li,[2] Jian-Min Yan,[4] Li Xie,[3] Meng Xu,[5] Lei Guo,[3] Shu-Juan Zhang,[2] Guan-Yin Gao,[6] Fei-Fei Wang,[7] Shan-Tao Zhang,[1] Xiaolin Wang,[8] Yang Chai,[4] Weiyao Zhao,[8,*] and Ren-Kui Zheng[3,9,*]

[1]National Laboratory of Solid State Microstructures, College of Engineering and Applied Science & Jiangsu Key Laboratory of Artificial Functional Materials & Collaborative Innovation Center of Advanced Microstructures, Nanjing University, Nanjing 210093, China

[2] School of Materials Science and Engineering and Jiangxi Engineering Laboratory for Advanced Functional Thin Films, Nanchang University, Nanchang 330031, P. R. China

[3] State Key Laboratory of High Performance Ceramics and Superfine Microstructure, Shanghai Institute of Ceramics, Chinese Academy of Sciences, Shanghai 200050, P. R. China

[4]Department of Applied Physics, The Hong Kong Polytechnic University, Hong Kong 999077, P. R. China

[5]College of Science, Hohai University, Nanjing 211189, P. R. China

[6]Hefei National Laboratory for Physical Sciences at the Microscale, University of Science and Technology of China, Hefei 230026, P. R. China

[7]Key Laboratory of Optoelectronic Material and Device, Department of Physics, Shanghai Normal University, Shanghai 200234, P. R. China

[8]Institute for Superconducting and Electronic Materials, & ARC Centre of Excellence in Future Low-Energy Electronics Technologies, Innovation Campus, University of Wollongong, NSW 2500, Australia

[9]School of Physics and Materials Science, Guangzhou University, Guangzhou 510006, P. R. China

*Correspondence to: weiyao.zhao@monash.edu (W.Y. Zhao) and zrk@ustc.edu (R.K. Zheng)

**ABSTRACT:** Spin-orbit interaction (SOI) offers a nonferromagnetic scheme to realize spin polarization through utilizing an electric field. Electrically tunable SOI through electrostatic gates have been investigated, however, the relatively weak and volatile tunability limit its practical applications in spintronics. Here, we demonstrate the nonvolatile electric-field control of SOI via constructing ferroelectric Rashba architectures, i.e., 2D $Bi_2O_2Se$/PMN-PT ferroelectric field effect transistors. The experimentally observed weak antilocalization (WAL) cusp in $Bi_2O_2Se$ films implies the Rashba-type SOI that arises from asymmetric confinement potential. Significantly, taking advantage of the switchable ferroelectric polarization, the WAL-to-weak localization (WL) transition trend reveals the competition between spin relaxation and dephasing process, and the variation of carrier density leads to a reversible and nonvolatile modulation of spin relaxation time and spin splitting energy of $Bi_2O_2Se$ films by this ferroelectric gating. Our work provides a scheme to achieve nonvolatile control of Rashba SOI with the utilization of ferroelectric remanent polarization.

## 1. Introduction

Spin-orbit interaction (SOI) has been intensively studied for related exotic physical phenomena and potential applications in spintronic devices and quantum computing fields [1-3]. SOI may result from either the lack of lattice inversion symmetry known as Dresselhaus effect, or Rashba type with broken structure symmetry imposed by external fields or asymmetric confinement potential [4,5]. Unlike traditional magnetic-field-induced spin polarization, SOI offers a nonferromagnetic scheme to realize the same purpose through utilizing an electric field.

And the effective modulation of SOI by electrostatic field have been explored in low-dimensional systems. However, the tunability of SOI through those dielectric gates are usually volatile and relatively weak, which is limited by the strength of electric field and low achievable areal charge densities $n_{2D} \sim 10^{12}$-$10^{13}$ cm$^{-2}$ [6-8].

Alternatively, ferroelectric (FE) materials with large spontaneous electric polarization can generate a strong local electric field, and modify the carrier densities of adjacent materials, thus leading to a greater tunability [9-11]. Moreover, ferroelectric gate can potentially tune the spin degree of freedom in a nonvolatile manner, which has been recently demonstrated in various spintronic devices, such as reversible and nonvolatile electrical switching of spin polarization and spin-charge conversion [12-16]. However, the nonvolatile manipulation of SOI involves replacing the dielectric gate by a ferroelectric one has been rarely reported so far. Thus, it is highly desired to realize a nonvolatile electric-field control of SOI using remanent polarization of ferroelectrics for the development of future spin memory and logic devices.

As an emerging 2D semiconductor material, the bismuth layered oxyselenide $Bi_2O_2Se$ is a good candidate for the modulation of SOI, which exhibits superior physical and chemical properties with a moderate band gap (0.8 eV), ultrahigh carrier mobility and good air stability [17-20]. Recently, some studies have reported excellent transport properties of $Bi_2O_2Se$, which stimulates further exploration of its novel quantum transport phenomena [19,21]. Therefore, it is of great significance to investigate the SOI in $Bi_2O_2Se$ system, which has been less experimentally studied until now. Moreover, the tetragonal crystal structure ($a = b = 3.89$ Å, $c = 12.21$ Å) of $Bi_2O_2Se$ enable it to be epitaxially grown on the widely used ferroelectric $Pb(Mg_{1/3}Nb_{2/3})O_3$-$PbTiO_3$ (PMN-PT, $a \sim b \sim c \sim 4.02$ Å) single crystal [Figure 1(a)]. And the areal carrier density of Bi$_2$O$_2$Se ($n_{2D}$ ~10$^{13}$ cm$^{-2}$) is lower than the $n_{2D}$ ~10$^{14}$ cm$^{-2}$ resulting from the ferroelectric remanent polarization ($P_r \approx$ 25-40 μC/cm$^2$) [10,19,22,23]. Hence, the 2D Bi$_2$O$_2$Se/PMN-PT heterostructure would be a potential candidate system, in which the SOI of Bi$_2$O$_2$Se could be reversibly and nonvolatilely manipulated via interfacial polarization charges of ferroelectric PMN-PT.

In this paper, we have achieved nonvolatile manipulation of SOI through constructing ferroelectric Rashba architectures, i.e., Bi$_2$O$_2$Se/PMN-PT 2D-FeFETs [Fig. 1(b)]. Different from tuning the interfacial spin-orbit coupling with ferroelectricity through the spin Hall angle, this work demonstrates the SOI tuning in Bi$_2$O$_2$Se/PMN-PT heterostructure by magnetotransport method. Upon switching the direction of ferroelectric polarization, the *in situ* electric-field control of carrier density, resistance and magnetoresistance are realized by the ferroelectric field effect. Further magnetotransport measurements demonstrate the presence of a Rashba-type SOI in Bi$_2$O$_2$Se films arising from the asymmetric confinement potential. The dependence of WAL to WL transition on polarization state, thickness, and temperature reveal that the crossover is determined by the relative scale of spin relaxation length and phase coherence length. Moreover, utilizing the ferroelectric gate, the SOI and spin-splitting energy could be modulated in a reversible and nonvolatile manner.

## 2. Experiments

Bi$_2$O$_2$Se films were deposited on (001)-oriented PMN-PT single-crystal substrates by the pulse laser deposition (PLD) with a XeCl excimer laser (λ = 308 nm). Before the deposition of films, the base pressure of the chamber was evacuated to a pressure lower than 5.0×10$^{-5}$ Pa. The working pressure was maintained at 3.0×10$^{-3}$ Pa during film deposition process. Other preparation parameters like the target-to-substrate distance, laser energy density, substrate temperature, and pulse repetition rate were kept at 6 cm, 2 J/cm$^2$, 400 ºC, and 3 Hz, respectively. After the completion of deposition, the as-grown films were cooled naturally to room temperature.

Ag electrodes with a thickness of 100 nm were deposited on the top surface of $Bi_2O_2Se$ films and the whole back of the bottom of PMN-PT substrates. The top Ag electrodes with different sizes are used to measure the electrical resistance and Hall resistance (Fig. S1 in the Supplemental Material [24]), respectively. The bottom Ag electrode together with one top Ag electrode was used to apply electric field to switch the polarization direction of the PMN-PT substrates.

The phase purity, out-of-plane and in-plane orientations, crystallinity of the grown films were analyzed by an x-ray diffractometer (PANalytical X'Pert PRO) equipped with CuK$\alpha_1$ radiation ($\lambda$=1.5406 Å). The crystal structure and interface epitaxy of the prepared heterostructures were also characterized using a transmission electron microscope (Tecnai G2 F20 S-Twin). The surface morphology of $Bi_2O_2Se$ films and piezoelectric properties of PMN-PT substrates were characterized via the atomic force microscopy (AFM) and piezoresponse force microcopy (PFM) using an atomic force microscope (MFP-3D, Asylum Research Inc.). The polarization-electric field (*P-E*) hysteresis loops of PMN-PT substrates were recorded by means of a TF3000 Ferroelectric Analyzer (aiXACCT, Germany) at room temperature. The electronic transport properties of $Bi_2O_2Se$ films including the resistance, magnetoresistance, and carrier density were measured via the standard four-probe technique and the Van der Pauw method, respectively, using a physical property measurement system (PPMS) (Quantum Design).

## 3. Results and discussion

The x-ray diffraction (XRD) $\theta$-$2\theta$ scan pattern in Fig. 1(c) shows that (00*l*)-oriented (*l* = 4, 6, 8, 10) diffraction peaks are detected for the $Bi_2O_2Se$ film, implying that the film is single phase and *c*-axis oriented. The XRD rocking curve taken on the Bi$_2$O$_2$Se (004) diffraction peak yields a full width at half maximum (FWHM) of 0.93º (Fig. S2 in the Supplemental Material [24]), indicating that the Bi$_2$O$_2$Se film has a relatively high degree of crystallinity. The homogeneous distribution of Bi, O, and Se elements (Fig. S3 in the Supplemental Material [24]) and a relatively smooth surface with a root-mean-square (RMS) roughness of 4.5 nm shown by AFM image (Fig. S4(a) in the Supplemental Material [24]) further suggest the good quality of the Bi$_2$O$_2$Se film. As depicted by the azimuthal $\phi$ scans in the inset of Fig. 1(c), the film and substrate both show a four-fold symmetry, which reveals that the Bi$_2$O$_2$Se film grows epitaxially on the PMN-PT substrate. In addition, the epitaxial relationship is further confirmed by the HRTEM image. As shown in Fig. 1(d), a thickness of a ~ 4 nm Bi$_2$O$_2$Se buffer layer is formed near the interface, and then the film grows layer by layer along the *c*-axis. The formation of the buffer layer is probably caused by the smaller lattice constant of Bi$_2$O$_2$Se (*a* ~ *b* ~ 0.39 nm) than that of PMN-PT (*a* ~ *b* ~ *c* ~ 0.40 nm), in order to release the tensile strain accumulated at the beginning of film deposition. Afterwards, the lattice spacing of 0.39 and 1.22 nm correspond to the lattice parameter *a* (*b*) and *c* of the Bi$_2$O$_2$Se unit cell, which manifests the well-defined atomic arrangement and epitaxial growth of the Bi$_2$O$_2$Se film.

The superior ferroelectric properties of the PMN-PT (001) substrate are illustrated by the complete the PFM images [Figs. S4(b)-4(d) in the Supplemental Material [24]). The shape of Δ*R*/*R* vs *E* curve in Fig. 1(e) is similar to that of the *P-E* loop of the PMN-PT [inset of Fig. 1(e)], implying that the ferroelectric field effect determines the resistance switching behaviors. Here, Δ*R*/*R* is defined as Δ*R*/*R*=[*R*(*E*)-*R*(0)]/*R*(0), where *R*(*E*) and *R*(0) are the resistance of the film with and without the application of external electric fields, respectively. After removing the electric field, a maximum value ∼ 30% of $\Delta R/R$ has been achieved for a 75-nm $Bi_2O_2Se$ film as the polarization state changes from $P_r^-$ to $P_r^+$ at room temperature, where $P_r^-$ and $P_r^+$ states represent the negative [top right inset in Fig. 1(f)] and positive [bottom left inset in Fig. 1(f)] polarization states of the PMN-PT substrate, respectively. Note that, the resistance relaxation behavior manifested by the non-square $\Delta R/R$ vs $E$ curve [Fig. 1(e)] could be caused by the trapping effects of defect states in the 4-nm disordered interfacial layer [25].

The negative Hall coefficients in Fig. 1(f) show that the 75-nm $Bi_2O_2Se$ film is an $n$-type semiconductor whose majority charge carriers are electrons. The correspondingly calculated room temperature carrier densities $n$ are $1.06\times10^{19}$ cm$^{-3}$ for the $P_r^+$ state and $0.81\times10^{19}$ cm$^{-3}$ for the $P_r^-$ state, respectively. In addition, the *in situ* XRD $\theta$-$2\theta$ scan in Fig. 1(g) proves that the polarization reversal of PMN-PT does not affect the strain state of the film since the diffraction peaks of the $Bi_2O_2Se$ film remain unchanged. All these evidences confirm that the resistance change of $Bi_2O_2Se$ film is dominated by the ferroelectric field effect. As schematically illustrated by the bottom left inset in Fig. 1(f), with the PMN-PT positively poled, the negative polarization charges are induced at the interface between $Bi_2O_2Se$ and PMN-PT, which will attract holes from the film to the interface region. This increases the electron carrier density of $Bi_2O_2Se$ film, leading to the decrease of the resistance of the film. The situation for the $P_r^-$ state is opposite by applying a negative electric field to PMN-PT (top right inset). Consequently, the nonvolatile and reversible modulation of resistance and carrier density have been realized in $Bi_2O_2Se$/PMN-PT 2D-FeFETs.

The low-temperature electronic transport properties of $Bi_2O_2Se$ films with several representative thicknesses are studied. With decreasing film thickness, the measured volume carrier density decreases monotonously as shown in Figure 2(a), and the discrepancy of the carrier density between the $P_r^+$ and $P_r^-$ states become more pronounced. This suggests the more prominent modulation ability of the ferroelectric field effect as the films become thinner, consistent with the largest Δ$R$/$R$ realized in the 75-nm film (Fig. S5 in the Supplemental Material [24]). On the other hand, the values of Ioffe-Regel parameter $k_F l = (3\pi^2 n)^{2/3} \hbar \mu / e$, where $\hbar$ is the reduced Planck constant and $\mu$ is the electron mobility (Fig. S6 in the Supplemental Material [24]), for different film thicknesses are obviously larger than one at low temperature ~ 200 K [Fig. 2(b)], implying that Bi$_2$O$_2$Se films are in the weakly disordered regime at this temperature range [26,27].

The magnetoresistance (MR) of the Bi$_2$O$_2$Se films in Fig. 2(c) exhibits a polarization-induced difference and show a similarly reduced magnitude of regulation with increasing film thickness. Here, MR is defined as MR = [$R(H)$-$R(0)$]/$R(0)$, $R(H)$ and $R(0)$ referring to the resistance in the presence and absence of a magnetic field $H$, respectively. As the film thickness increases, in the range of $H$ > 1 T, the shape of MR varies from a parabola pointing downwards, to a double-dip W-shaped curve, finally to a parabola pointing upwards, along with the sign changing from negative to positive. This transition behavior of MR including the sign and magnitude is strongly influenced by different carrier density [28,29]. On the other hand, the MR of the 75-nm Bi$_2$O$_2$Se film with increasing temperature in Fig. 2(d) and Figs. S7-S9 in the Supplemental Material [24] resembles the curve-shape transitions mentioned above. Particularly, a sharp cusp near zero-magnetic-field in the MR curve is observed at low temperatures. This is a characteristic signature of weak antilocalization (WAL) effect, which implies the presence of spin-orbit interaction (SOI) in the Bi$_2$O$_2$Se films.

As an effective method to identify SOI, the low-field magnetoconductance behavior due to the weak localization (WL) and WAL have been extensively employed to investigate spin relaxation processes in various systems [30]. For a 2D disordered system, the low-field WAL effect with different mechanisms can be described by the Iordanskii-Lyanda-Geller-Pikus (ILP) theory (Equation S1) or the Hikami-Larkin-Nagaoka (HLN) theory [31,32]. As the assumption is satisfied from the Ioffe-Regel parameter $k_F l > 1$, the fittings of low-field magnetoconductance $\Delta G = G(H)-G(0) = R^{-1}(H)-R^{-1}(0)$ with both theories are shown in Figure 3. It is clear that the HLN fitting provides better agreement with the experimental data (Table S1), proving that the HLN theory is more effective to describe WAL in the $Bi_2O_2Se$ film, consistent with the earlier literature [33]. Specially, the magnetoconductance of the WAL effect using HLN theory is given as below:

$$\Delta G(B) = -\frac{e^2}{\pi h}\left[\frac{1}{2}\psi\left(\frac{B_\varphi}{B}+\frac{1}{2}\right)-\frac{1}{2}\ln\left(\frac{B_\varphi}{B}\right)+\psi\left(\frac{B_{so}+B_{el}}{B}+\frac{1}{2}\right)-\ln\left(\frac{B_{so}+B_{el}}{B}\right)\right.$$
$$\left.-\frac{3}{2}\psi\left(\frac{(4/3)B_{so}+B_\varphi}{B}+\frac{1}{2}\right)+\frac{3}{2}\ln\left(\frac{(4/3)B_{so}+B_\varphi}{B}\right)\right] \quad (1)$$

Here, $\Psi$ is the digamma function. Three subscripts $\varphi$, $so$ and $el$ denote inelastic dephasing process, spin-orbit scattering, and elastic scattering, respectively. The characteristic magnetic field $B_\varphi$, $B_{so}$ and $B_{el}$ could be extracted from the fitting to the experimental data. The mean free path $L_{el} = (\hbar/4eB_{el})^{1/2}$ (Fig. S12 in the Supplemental Material [24]) are much smaller than the distance between two top-top electrodes, and thus the carrier transport in $Bi_2O_2Se$ films is in the diffusive regime [33].

In order to determine the strength of SOI, the low-field $\Delta G$ of the 75-nm $Bi_2O_2Se$ film for the $P_r^+$ and $P_r^-$ states are fitted and plotted in Figure 4(a) for the most notable WAL feature, and other $\Delta G$ data of films with different thicknesses are shown in (Figs. S10-S11 in the Supplemental Material [24]). At $T=3$ K, the observed sharp WAL peak of $\Delta G$ near zero field is caused by the positive correction to the conductance with the presence of SOI. With heating to 30 K, the sharp peak is gradually suppressed and eventually develops into a broad dip, a signature of WL effect. Hence, the temperature dependent WAL-WL transition is observed here, as reported in previous trilayer Graphene, SrTiO$_3$-FETs, and InAs/GaSb double quantum well [34-36]. Intriguingly, upon the ferroelectric polarization switching from the $P_r^+$ state to the $P_r^-$ state of PMN-PT substrate, the WAL effect is suppressed as reflected by the decrease in the magnitude of $\Delta G$ and a broaden of the WAL peak of the Bi$_2$O$_2$Se film, exhibiting a polarization-induced tendency from WAL to WL.

To investigate the spin relaxation and phase coherence process, temperature-dependent spin relaxation length $L_{so}$ and dephasing length $L_\varphi$ calculated from $L_{so,\varphi} = (\hbar/4eB_{so,\varphi})^{1/2}$ are plotted in Figs. 4(b) and 4(d), respectively. The $L_\varphi$ show almost no dependence on the polarization states [Figs. 4(b)], and basically follow a power law of $L_\varphi \sim T^{-0.5}$ [Figs. 4(c) and 4(e)]. This indicates that the dephasing process is primarily owing to the Nyquist scattering, i.e., electron-electron scattering with small energy transfers [37,38]. In contrast, the $L_{so}$ presents a clear polarization-induced difference [Fig. 4(d)], which are attributed to the modulation of the carrier density by the ferroelectric field effect. The values of $L_{so} \sim 84$ nm for the $P_r^+$ state and $\sim 102$ nm for the $P_r^-$ state at 3 K, shorter than $\sim 290$ nm in Al$_x$Ga$_{1-x}$N/GaN 2DEG and $\sim 250$ nm in InSb nanowires, prove the presence of strong SOI in the Bi$_2$O$_2$Se film [39,40]. Here, because of the inherent inversion symmetry presented in Bi$_2$O$_2$Se, and given the electric-field induced polarization charges, therefore, the Rashba-type SOI is the most likely dominant mechanism in the 2D Bi$_2$O$_2$Se film, consistent with previous reports [41,42]. Remarkably, the polarization induced controllability of $L_{so}$ is much stronger than the conventional dielectric gating effect in Bi$_2$O$_2$Se nanoplates [33]. These results significantly demonstrate ferroelectric gating control of the SOI in the Bi$_2$O$_2$Se/PMN-PT system.

Furthermore, $L_{so}$ and $L_\varphi$ for the $P_r^+$ and $P_r^-$ states are respectively plotted in Figs. 4(c) and 4(e), and show a crossover with increasing temperature. This nicely illustrates the key aspect that when the $L_{so}$ is shorter than $L_\varphi$ for $T<15$ K, the WAL is observed because the spin precession destroys the phase coherence of electron wave function on time-reversed closed path; otherwise, the WL is dominant due to thermally enhanced dephasing process [43]. As the carrier density for the $P_r^+$ state is higher than that for the $P_r^-$ state for the 75-nm film [Fig. 2(a)], this results in shorter $L_{so}$ and larger $L_\varphi/L_{so}$ for the $P_r^+$ state, which gives rise to the stronger WAL, as manifested by the sharper negative $\Delta G$ in Fig. 4(a). Consequently, the temperature corresponding to the intersection point between $L_{so}$ and $L_\varphi$ for the $P_r^+$ state is larger than that for the $P_r^-$ state [Fig. 4(c) and 4(e)], consistent with the transition trend from WAL to WL in the $\Delta G$ curves. All these analyses fully prove that the WAL effect and Rashba-type SOI in Bi$_2$O$_2$Se film could be nonvolatilely manipulated by switching the ferroelectric polarization.

Thickness-dependent magnetoconductance $\Delta G$ of the Bi$_2$O$_2$Se films with thicknesses ranging from 75 to 150 nm at 3 K are shown in Figure 5(a). All theoretical fittings with HLN theory to the experimental data are in good agreement. As the films become thicker, the characteristic WAL peak around zero magnetic field is gradually weakened, presenting a thickness-dependent WAL-WL crossover. Moreover, the derived spin relaxation length $L_{so}$, dephasing length $L_\varphi$ and mean free path $L_{el}$ basically decrease with increasing film thickness (Fig. S12 in the Supplemental Material [24]). As a result, $L_\varphi$ gradually approaches to $L_{so}$ in thicker films, leading to neglectable WAL effect in the 150-nm film [Fig. 5(a)]. And this is also reflected by the intersection point between $L_{so}$ and $L_\varphi$ shifting to lower temperature with increasing film thickness [Figs. 4(c) and 4(e) and (Figs. S10 and S11 in the Supplemental Material [24])]. Thus, the stronger WAL effect and more effective manipulation of SOI through ferroelectric field effect is expected in thinner $Bi_2O_2Se$ film whose carrier density could be more effectively modulated by ferroelectric polarization.

Significantly, the influences of the ferroelectric field effect on the phase and spin relaxation times are further elaborated. The phase relaxation time can be obtained by $\tau_\varphi = \hbar/(4eDB_\varphi)$, where the diffusion coefficient $D = v_F^2 \tau_{tr}/2$ is calculated by the Fermi velocity $v_F = k_F \hbar/m^*$ and transport relaxation time $\tau_{tr} = \mu m^*/e$ with Fermi wave vector $k_F = (3\pi n)^{1/3}$ and effective electron mass $m^* = 0.184 m_e$. As shown by Fig. 5(b), the temperature dependence of $\tau_\varphi$ scales as $\tau_\varphi \sim T^{-1}$, which could be evidenced again that the Nyquist dephasing with electron-electron interaction is the main phase relaxation mechanism [37]. However, for thicker $Bi_2O_2Se$ films, the variation trend of $\tau_\varphi$ vs $T$ diverges from the dotted line $\sim T^{-1}$, which may originate from the deviation from the 2D nature. On the other hand, the spin relaxation time $\tau_{so}$ is given by $\tau_{so} = \hbar/(4eDB_{so})$. Taking the 75-nm film as an example, this yields a spin relaxation time $\tau_{so} \sim 9$ ps for the $P_r^-$ state, much larger than $\tau_{so} \sim 4$ ps for the $P_r^+$ state (Table 1), due to the stronger SOI in the higher carrier density state [33,37]. Again, this result demonstrates the polarization-induced tunability of SOI upon switching the polarization state from $P_r^-$ to $P_r^+$.

**Table 1.** The values of transport parameters and SOI related parameters for 75-nm film at 3 K.

| | $n$ (cm$^{-3}$) | $\mu$ (cm$^2$ V$^{-1}$ s$^{-1}$) | $D$ (m$^2$/s) | $L_{so}$ (nm) | $L_\varphi$ (nm) | $\tau_{so}$ (ps) | $\Delta_{so}$ (meV) | $\beta$ (eV Å$^3$) |
|---|---|---|---|---|---|---|---|---|
| $P_r^+$ | 1.56×10$^{19}$ | 133.33 | 0.0017 | 84.0 | 186.4 | 4.29 | 1.91 | 2.07 |
| $P_r^-$ | 1.07×10$^{19}$ | 115.95 | 0.0011 | 102.2 | 195.4 | 9.35 | 1.38 | 2.18 |

Given that Rashba SOI contributes to band spin spitting, the corresponding spin splitting energy is calculated using $\Delta_{so} = \hbar/(2\tau_{tr}\tau_{so})^{1/2}$. Carrier density dependent $\Delta_{so}$ plotted in Fig. 5(d) shows that $\Delta_{so}$ increases linearly with the carrier density for $Bi_2O_2Se$ films with different thicknesses. This also supports that the dominant SOI mechanism is the Rashba-type [36]. Therefore, the Rashba coefficient $\beta$ for SOI could be further obtained from $4\beta^2 k_F^6 = \Delta_{so}^2 = \frac{2e\hbar D}{\tau_{tr}} B_{so}$. The calculated value of $\beta$ presents a systematic variation trend in relation to the carrier density [Fig. 5(e)], similar to some systems like 2DEG in $SrTiO_3$-FETs, 2D hole gas in gated diamond devices [35,38]. Notably, the values of $\Delta_{so}$ derived here, e.g., 1.3-1.9 meV for the 75-nm $Bi_2O_2Se$ film (Table 1), exceed many other 2D systems in terms of magnitude [44,45]. More impressively, the nonvolatile modulation of $\Delta_{so}$ realized here through ferroelectric polarization is intrinsically different from those volatile one after removing electric field in dielectric gates.

## Conclusions

In present work, the 2D $Bi_2O_2Se$ films are epitaxially grown on ferroelectric PMN-PT (001) substrates to form $Bi_2O_2Se$/PMN-PT 2D-FeFETs. The *in situ* electric-field control of the carrier density, resistance, and magnetoresistance are realized by ferroelectric field effect. Further, the polarization state, film thickness, and temperature dependent WAL effect are systematically measured and analyzed. The low-field magnetoconductance of WAL with HLN fittings demonstrate the presence of a strong spin-orbit interaction in 2D $Bi_2O_2Se$ films due to the asymmetric confinement potential. The crossover of WAL to WL is primarily determined by the relative scale of spin relaxation length and dephasing length, and the main phase relaxation mechanism is the Nyquist scattering. Remarkably, the modulation of spin-orbit interaction realized by ferroelectric gating is reversible and nonvolatile. Our work manifests that heterostructure systems utilizing films with Rashba SOI and ferroelectric polarization could achieve reversible and nonvolatile manipulation of SOI, which may inspire schemes for spintronics and related quantum applications.

## Acknowledgements

This work was supported by the National Natural Science Foundation of China (Grant No. 11974155). X.W. and W.Z. acknowledge the supporting form ARC Centre of Excellence in Future Low-Energy Electronics Technologies CE170100039.

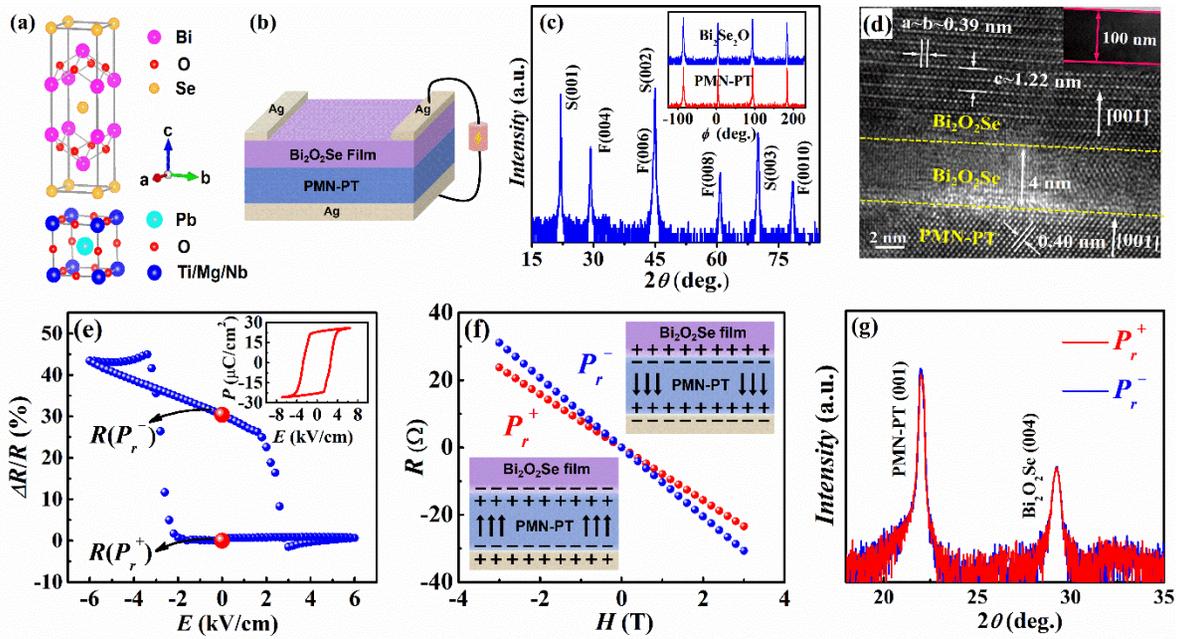

FIG. 1. (a) Crystal structure of tetragonal $Bi_2O_2Se$ and pseudocubic PMN-PT. (b) Schematic diagram of the $Bi_2O_2Se$/PMN-PT 2D-FET device. (c) XRD $\theta$-$2\theta$ scan and $\phi$ scan patterns (inset) for the $Bi_2O_2Se$/PMN-PT heterostructure. Here, F and S represent the $Bi_2O_2Se$ film and PMN-PT substrate, respectively. (d) TEM image taken on the interface of a 100-nm $Bi_2O_2Se$/PMN-PT. (e) The relative resistance change ($\Delta R/R$) of the 75-nm $Bi_2O_2Se$ film as a function of electric field $E$ applied to PMN-PT, the inset showing ferroelectric polarization versus electric field (P-E) loop of PMN-PT (001). (f) Room temperature Hall resistance of the 75-nm $Bi_2O_2Se$ film versus magnetic field when PMN-PT was in the positive ($P_r^+$) and negative ($P_r^-$) poled states, respectively, and schematic illustration for the accumulation of positive (top right inset) and negative (bottom left inset) polarization charges under $P_r^+$ and $P_r^-$ states of PMN-PT. (g) XRD $\theta$-$2\theta$ scan pattern of the $Bi_2O_2Se$ film for the $P_r^+$ and $P_r^-$ states of PMN-PT.

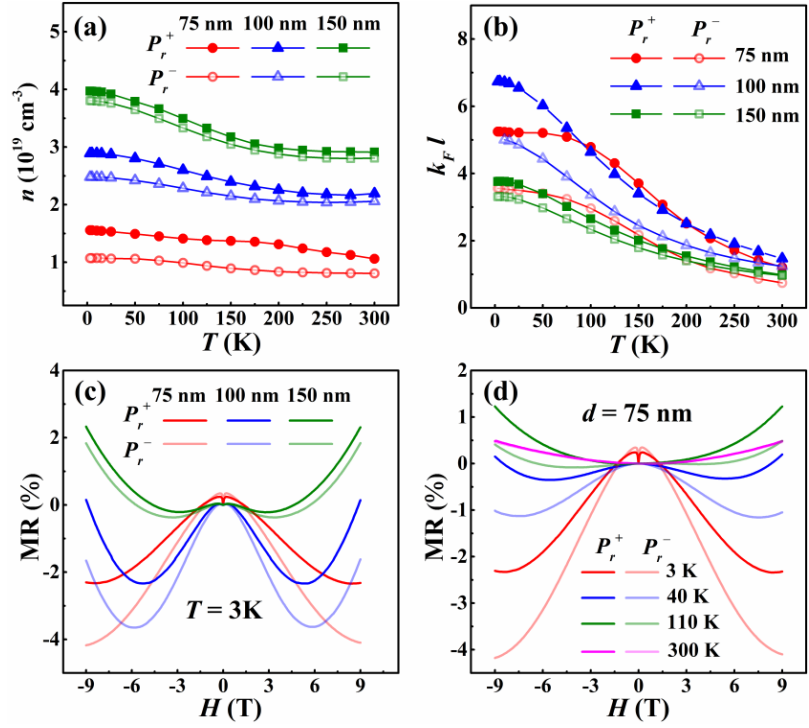

FIG. 2. (a) Temperature dependence of the carrier density of $Bi_2O_2Se$ films with different thicknesses. (b) Temperature dependence of the Ioffe-Regel parameter ($k_Fl$) of $Bi_2O_2Se$ films with different thicknesses. (c) Magnetic field dependent magnetoresistance (MR) of the $Bi_2O_2Se$ films with different thicknesses at 3 K. (d) Magnetic field dependent MR of the 75-nm $Bi_2O_2Se$ film at various fixed temperatures.

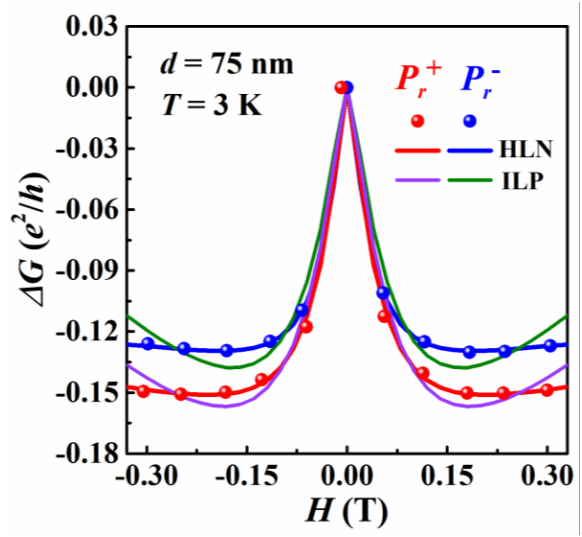

FIG. 3. Experimental WAL data of the 75-nm $Bi_2O_2Se$ film for the $P_r^+$ and $P_r^-$ states of PMN-PT, as measured at $T =3$ K. The solid lines are fitted with different theories (red and blue line for the HLN theory, purple and green line for the ILP theory).

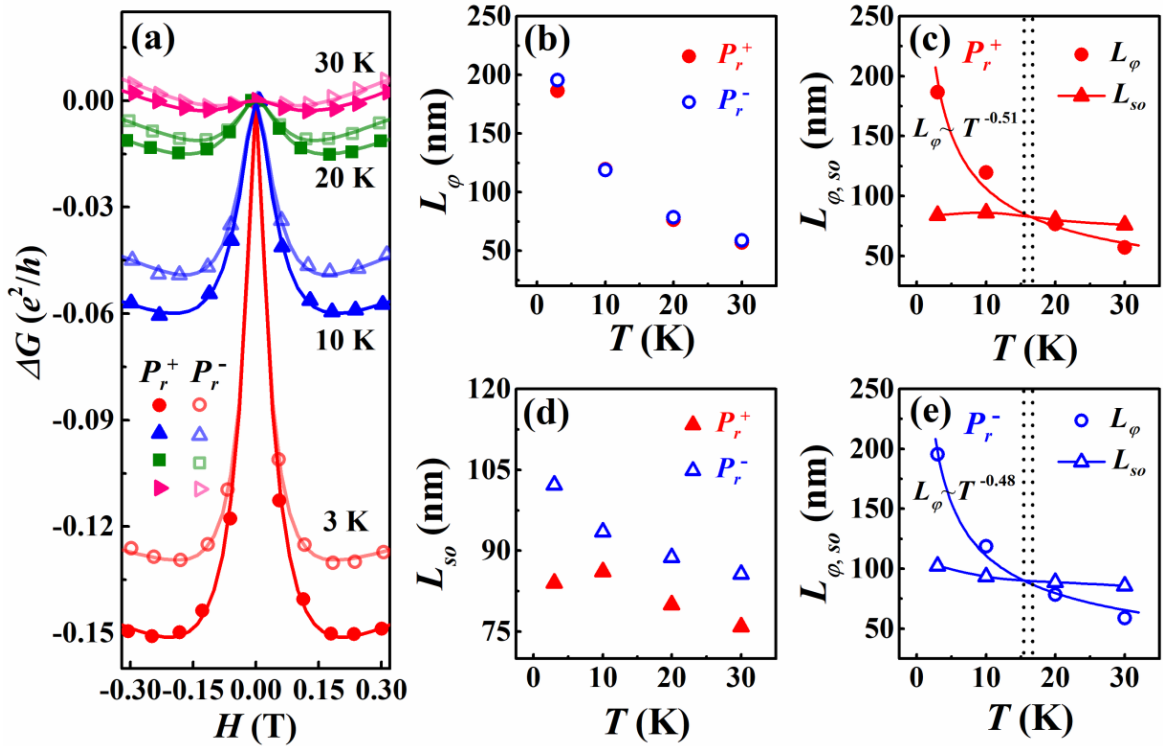

FIG. 4. (a) Low-field magnetoconductance $\Delta G$ ($e^2/h$) of the 75-nm $Bi_2O_2Se$ film for the $P_r^+$ and $P_r^-$ states at different fixed temperatures. The solid lines represent the fitting results obtained with HLN equation. (b) Dephasing length ($L_\varphi$) and (d) spin relaxation length ($L_{so}$) as a function of temperature, extracted from the $\Delta G(e^2/h)$ data for the $P_r^+$ and $P_r^-$ states. Characteristic $L_{so}$ and $L_\varphi$ involved in WAL and WL as a function of temperature for (c) $P_r^+$ and (e) $P_r^-$ states, and specially, the solid lines for $L_\varphi$ in (c) and (e) are the fits to experimental data by power function.

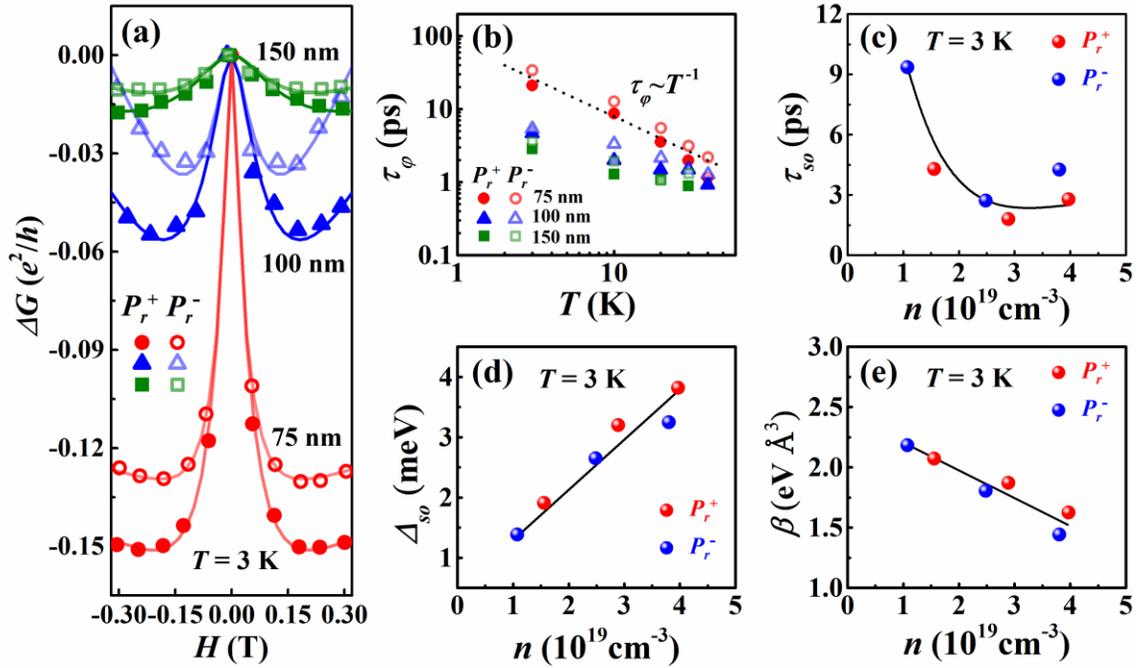

FIG. 5. (a) Magnetoconductance of Bi$_2$O$_2$Se films with different thicknesses for the $P_r^+$ and $P_r^-$ states of PMN-PT. The solid lines show fittings of experimental data with HLN equation. (b) Temperature dependence of phase relaxation time $\tau_\varphi$ of Bi$_2$O$_2$Se films with different thicknesses, the dotted line to signify $\tau_\varphi \sim T^{-1}$ for Nyquist

dephasing. (c) Spin relaxation time $\tau_{so}$ as a function of carrier density, the line is a guide to the eye. (d) Variation of the spin-orbit splitting energy $\Delta_{so}$ with the volume carrier density $n$. (e) Carrier density dependent Rashba coefficient $\beta$. The solid lines in (d) and (e) are linear fits to the data.